\documentclass[10pt,twocolumn]{article}

\usepackage{listings}
\usepackage{makeidx}
\usepackage{algorithm,algpseudocode}
\usepackage{graphicx}
\usepackage{float}
\usepackage{cite}
\usepackage{amsmath}
\usepackage{rotating,todonotes, xspace}
\usepackage{multirow}
\usepackage{hyperref}
\usepackage{xspace}
\usepackage{paralist}
\usepackage{booktabs}
\usepackage{balance}
\usepackage{cleveref}
\usepackage{color, colortbl}

\definecolor{Gray}{gray}{0.9}
\newcommand{\pp}[1]{\vspace{6pt}\noindent\textbf{\emph{#1.}}\xspace}

\begin{document}

\title{WorkflowHub: Community Framework for Enabling Scientific Workflow Research and Development -- Technical Report}

\author{
  \normalsize
  Rafael Ferreira da Silva$^1$,
  Lo\"ic Pottier$^1$,
  Tain\~a Coleman$^1$,
  Ewa Deelman$^1$,
  Henri Casanova$^2$
  \\
  \normalsize
  $^1$University of Southern California, $^2$University of Hawai'i at Mano\=a
}

\date{}

\maketitle

\begin{abstract}
Scientific workflows are a cornerstone of modern scientific computing. They
are used to describe complex computational applications that require
efficient and robust management of large volumes of data, which are
typically stored/processed at heterogeneous, distributed resources. The
workflow research and development community has employed a number of
methods for the quantitative evaluation of existing and novel workflow
algorithms and systems. In particular, a common approach is to simulate
workflow executions.
In previous
work, we have presented a collection of tools that have been used for
aiding research and development activities in the  Pegasus project, and
that have been adopted by others for conducting workflow research.  Despite
their popularity, there are several shortcomings that prevent easy
adoption, maintenance, and consistency with the evolving structures and
computational requirements of production workflows.  In this work, we
present \emph{WorkflowHub}, a community framework that provides a
collection of tools for analyzing workflow execution traces, producing
realistic synthetic workflow traces, and simulating workflow executions.
We demonstrate the realism of the generated synthetic traces by comparing
simulated executions of these traces with actual workflow executions.
We also contrast these results with those obtained when using the previously
available
collection of tools. We find that our framework not only can be used to
generate representative synthetic workflow traces (i.e., with workflow
structures and task characteristics distributions that resembles those in
traces obtained from real-world workflow executions), but can also generate
representative workflow traces at larger scales than that of available workflow
traces.

\end{abstract}


\section{Introduction}
\label{sec:introduction}

Scientific workflows are relied upon by thousands of
researchers~\cite{deelman2019evolution} for managing data analyses,
simulations, and other computations in almost every scientific
domain~\cite{liew2016scientific}. Scientific workflows have underpinned
some of the most significant discoveries of the last
decade~\cite{deelman-fgcs-2015, klimentov2015next}. These discoveries are
in part a result of decades of workflow management system (WMS) research,
development, and community engagement to support the
sciences~\cite{osti_1422587}.  As workflows continue to be adopted by
scientific projects and user communities, they are becoming more complex
and require more sophisticated workflow management capabilities. Workflows
are being designed that can analyze terabyte-scale datasets, be composed of
millions of individual tasks that execute for  milliseconds up to several
hours, process data streams, and process static data in object stores.
Catering to these workflow features and demands requires WMS research
and development at several levels, from algorithms and systems all the way
to user interfaces.

A traditional approach for testing, evaluating, and evolving WMS is to use
full-fledged software stacks to execute applications on distributed platforms
and testbeds. Although seemingly natural, this approach has severe shortcomings
including lack of reproducible results, limited platform configurations, and
time and operational costs. An alternative that does not have these
shortcoming is to use simulation, i.e.,
implement and use a software artifact that models the functional and
performance behaviors of software and hardware stacks of interest. Thus,
the scientific workflow community has leveraged simulation for the development
and evaluation of, for example, novel algorithms for scheduling, resource
provisioning, and energy-efficiency, workflow data footprint constraints,
exploration of data allocation strategies, among others~\cite{canon2020,han2019}.

Studying the execution of workflows in simulation requires sets of
workflow applications to be used as benchmarks. This is so that quantitative
results are obtained for a range of representative workflows.
In~\cite{ferreiradasilva-escience-2014}, we have described a collection
of tools and data that together have enabled research and
development of the Pegasus~\cite{deelman-fgcs-2015} WMS. These
community resources have enabled over 30 research papers\footnote{Based on
records provided by Google Scholar.} by providing synthetic workflow
traces for evaluation via simulation. Despite the extensive usage of this
pioneer effort, it lacks (\emph{i})~a common format for
representing workflow execution traces in a way that is agnostic to workflow systems;
(\emph{ii})~structured methods for encoding the workflow design and structure;
(\emph{iii})~solid techniques for generating synthetic workflows in which
workflow characteristics conform to the original workflow features; and
(\emph{iv})~a set of tools for analyzing workflow traces,
which would support the integration of traces from new application domains.

In this paper, we present the \emph{\textbf{WorkflowHub
project}}~\cite{workflowhub}, an open source community framework that
provides a collection of structured methods and techniques, implemented as
part of usable tools, for analyzing workflow traces and producing
synthetic, yet realistic, workflow traces. WorkflowHub mitigates the
shortcomings of our previous set of tools by using a common JSON format for
representing workflow traces.
Any workflow execution log can be captured into
this system-agnostic format.
In addition, WorkflowHub provides an open source Python package to analyze traces and generate representative synthetic traces in that same format.
Workflow simulators that support this format can then take
real-world and synthetic workflow traces as input for driving the simulation.
Figure~\ref{fig:concept} shows an overview of the
WorkflowHub conceptual architecture. Information  in workflow execution logs
is extracted as \textbf{workflow traces} using the common JSON format.
Workflow ``recipes" are obtained from the analysis of such traces.
More precisely, these recipes embody results from statistical analysis and
distribution fitting performed for each workflow task type so as to
characterize task runtime and input/output data sizes.  These recipes are
then used for informing a \textbf{workflow generator}, which produces
synthetic representative workflow traces. Finally, these traces can be used by a
\textbf{workflow simulator} for conducting experimental workflow research
and development.  Specifically, this work makes the following
contributions:

\begin{figure}[!t]
  \centering
  \includegraphics[width=\linewidth]{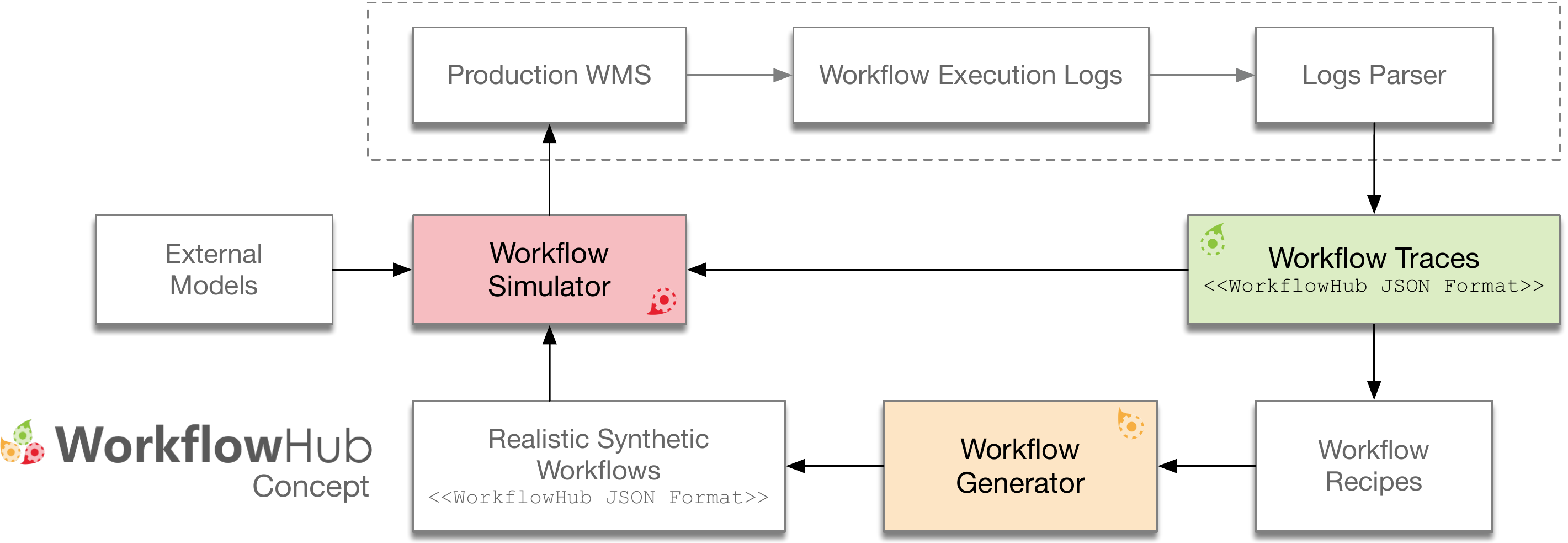}
  \caption{The WorkflowHub conceptual architecture.}
  \label{fig:concept}
\end{figure}

\begin{compactenum}
  \item A collection of workflow execution traces acquired from actual
        executions of state-of-the-art compute- and data-intensive workflows
        in a cloud environment;
  \item A common format for representing both collected traces and
        generated synthetic traces;
  \item An open source Python package~\cite{workflowhub-python} that provides
        methods for analyzing traces, deriving recipes, and generating
        representative synthetic traces;
  \item A collection of open-source workflow management systems simulators
        and simulation frameworks that support our common format;
  \item An evaluation of the accuracy of WorkflowHub's generated workflows
        at scale via a simulation case study, and a comparison to our previous
        set of tools~\cite{ferreiradasilva-escience-2014}.
\end{compactenum}

\section{Related Work}
\label{sec:related-work}

Workload archives are widely used for distributed computing research, to
validate assumptions, to derive workload models, and to perform experiments
by replaying workload executions, either in simulation or on real-world platforms.
Available repositories, such as the Parallel Workloads
Archive~\cite{feitelson2014experience}, the Grid Workloads
Archive~\cite{iosup2008grid}, and the Failure Trace
Archive~\cite{kondo2010failure}, contain data acquired at the
infrastructure level at compute sites, or by monitoring or obtaining logs
from deployed compute services. The workloads in these archives do include
data generated by workflow executions. But the information captured is
about individual job executions and about resource utilization. As
a result, there is at best little information on the task dependency
structure of workflows.

In the context of scientific workflows, the myExperiment
project~\cite{goble2010myexperiment} enables users to share workflows and
their semantics, however it does not collect metrics from
workflow executions. The Common Workflow Language
(CWL)~\cite{amstutz2016common} is an open standard for describing workflows
in a way that makes them portable and scalable across a variety
of software and hardware environments. Our proposed common format
(described below) is mostly inspired by the CWL standard, though our format
captures performance metrics data (e.g., total amount of I/O bytes read and
written, power consumption, etc.) and compute resource characteristics,
which are key for generating realistic workflow traces. The recently
established Workflow Trace Archive~\cite{versluis2020workflow} is an
open-access archive that provides a collection of execution traces from
diverse computing infrastructures and tools for parsing, validating, and
analyzing traces. To date, the archive has collected traces from 11
existing online repositories (including 10 traces obtained
from a preliminary version of WorkflowHub) and uses an object-oriented
representation (using the Parquet columnar storage format for Hadoop) for
documenting traces. Our common format instead uses JSON, which provides
easier time mapping to domain objects, regardless of the programming
language used for processing traces.
Also, the format used in~\cite{versluis2020workflow} captures workflow
executions information in terms of resource usage on the specific hardware
platform used to execute the workflow. As a result, it is difficult to use
this information to reconstruct a platform-independent, abstract workflow
structure.  By contrast, while WorkflowHub also records platform-specific
behaviors in its traces, in addition it ensures that the abstract workflow
structure is directly available from these traces.  This is crucial for
research purposes, as abstract workflow structures are needed for,
for instance, simulating workflow executions  on platform configurations that
differ from that used to collect the workflow execution trace.

Several studies have used synthetic workflows to explore how different
workflow features impact execution and interplay with each other (e.g.,
number of tasks, task dependency structure, task execution times).  Tools
such as SDAG~\cite{amer2012evaluating} and
DAGEN-A~\cite{amalarethinam2011dagen} generate random synthetic workflows,
but these are  not necessarily representative of real-world scientific workflows.
In our previous work~\cite{ferreiradasilva-escience-2014}, we developed a
tool for generating synthetic workflow configurations based on real-world
workflow  instances. As a result, the overall structure of generated
workflows was reasonably representative of real-world workflows. But that
tool uses only two types of statistical distributions  (uniform and
normal), and as a result workflow performance behavior may not be
representative (see results in Section~\ref{sec:experiments}).

\section{The WorkflowHub}
\label{sec:workflowhub}

The WorkflowHub project (\url{https://workflowhub.org}) is a community
framework for enabling scientific workflow research and development. It
provides foundational tools for analyzing workflow execution traces, and
generating synthetic, yet realistic, workflow traces. These traces can then
be used for experimental evaluation and development of novel algorithms
and systems for overcoming the challenge of efficient and robust
execution of ever-demanding workflows on increasingly complex distributed
infrastructures.

Figure~\ref{fig:concept} shows an overview of the workflow research life cycle
process that integrates the three axes of the WorkflowHub project:
(\emph{i})~workflow execution traces, (\emph{ii})~workflow generator, and
(\emph{iii})~workflow simulator.

\subsection{Workflow Execution Traces}
\label{sec:traces}

\begin{table*}[!t]
  \centering
  \scriptsize
  \setlength{\tabcolsep}{4pt}
  \begin{tabular}{lllrrp{6cm}}
    \toprule
    Application & Science Domain & Category & \# Traces & \# Tasks & Runtime and Input/Output Data Sizes Distributions \\
    \midrule
    1000Genome  & Bioinformatics  & Data-intensive    & 22 &  8,844
                & alpha, chi2, fisk, levy, skewnorm, trapz \\
    \rowcolor{Gray}
    Cycles      & Agroecosystem   & Compute-intensive & 24 & 30,720
                & alpha, beta, chi, chi2, cosine, fisk, levy, pareto,
                  rdist, skewnorm, triang \\
    Epigenomics & Bioinformatics  & Data-intensive    & 26 & 15,242
                & alpha, beta, chi2, fisk, levy, trapz, wald \\
    \rowcolor{Gray}
    Montage     & Astronomy       & Compute-intensive &  8 & 32,606
                & alpha, beta, chi, chi2, cosine, fisk, levy, pareto,
                  rdist, skewnorm, wald \\
    Seismology  & Seismology      & Data-intensive    & 11 &  6,611
                & alpha, argus, fisk, levy \\
    \rowcolor{Gray}
    SoyKB       & Bioinformatics  & Data-intensive    & 10 &  3,360
                & argus, dweibull, fisk, gamma, levy, rayleigh, skewnorm,
                  triang, trapz, uniform \\
    \midrule
    \textbf{6 applications} & \textbf{4 domains} & \textbf{2 categories} & \textbf{101} & \textbf{97,383} & \textbf{18 probability distributions} \\
    \bottomrule
  \end{tabular}
  \caption{Collection of workflow execution traces hosted by WorkflowHub. All traces were obtained using the Pegasus WMS running on the Chameleon cloud platform.}
  \label{tab:traces}
\end{table*}

The first axis of the WorkflowHub project targets the
collection and curation of open access production workflow execution traces
from various scientific applications, all made available using a common trace format.
A workflow execution trace is built based on logs of an actual execution of a scientific
workflow on a distributed platform (e.g., clouds, grids, clusters).
More specifically,
the three main types of information included in the trace are:

\begin{compactitem}
\item workflow task execution metrics (runtime, input and output data sizes,
      memory used, energy consumed, CPU utilization, compute resource that
      was used to execute the task, etc.);
\item workflow structure information (inter-task control and data dependencies); and
\item compute resource characteristics (CPU speed, available RAM, etc.).
\end{compactitem}

\pp{The WorkflowHub JSON format}
The WorkflowHub project uses a common format for representing collected workflow
traces and generated synthetic workflows traces. Workflow
simulators and simulation frameworks that support this common
format can then use both types of traces interchangeably. This common format uses a JSON
specification (publicly available on GitHub~\cite{workflowhub-schema}), which
captures all relevant trace information as listed above.
The GitHub repository also provides a Python-based
JSON schema validator for verifying the syntax of JSON trace files, as well
as their semantics, e.g., whether all files and task dependencies are
consistent. Users are encouraged to contribute additional workflow traces
for any scientific domain,
as long as they conform to the WorkflowHub's common format.

\pp{Collection of traces}
An integral objective of the WorkflowHub project is to collect and
reference open access workflow traces from production workflow systems.
Table~\ref{tab:traces} summarizes the set of workflow traces currently
hosted on WorkflowHub. These traces are from six representative science
domain applications, in which workflows are composed of compute- and/or
data-intensive tasks. (Note that although a workflow may be categorized
overall as, for example, data-intensive, it may be composed of
different kinds of tasks including, e.g., CPU-intensive ones.)  We argue that the 101
archived workflow traces form a representative set of small- and
large-scale workflow configurations. In addition to consuming/producing large
volumes of data processed by thousands of compute tasks, the structures of
these workflows are sufficiently complex and heterogeneous to encompass
current and emerging large-scale workflow execution
models~\cite{ferreiradasilva-fgcs-2017}.

\subsection{Workflow Trace Generator}
\label{sec:generator}

Workflow execution traces are commonly used to drive experiments for
evaluating novel workflow algorithms and systems. It is crucial to run
large numbers of such experiments for many different  workflow
configurations, so as to ensure generality of obtained results.  In
addition, it is useful to conduct experiments while varying one or more
characteristics of the workflow application, so as to study how these
characteristics impact workflow execution. For instance, one may wish, for a
particular overall workflow structure, to study how the workflow execution
scales as the number of tasks increases.  And yet, current archives only
include traces for limited workflow configurations. And even as efforts are
underway, including WorkflowHub, to increase the size of these archives, it
is not realistic to expect them to include all relevant workflow
configurations for all experimental endeavors. Instead, tools must be
provided to generate representative \emph{synthetic} workflow traces. These
traces should be generated based on real workflow traces, so as to be
representative, while conforming to user-specified characteristics, so as to be useful.
The second axis of the WorkflowHub project targets the generation of such
realistic synthetic workflow traces with a variety of characteristics.

\begin{figure}[!t]
  \centering
  \includegraphics[width=0.7\linewidth]{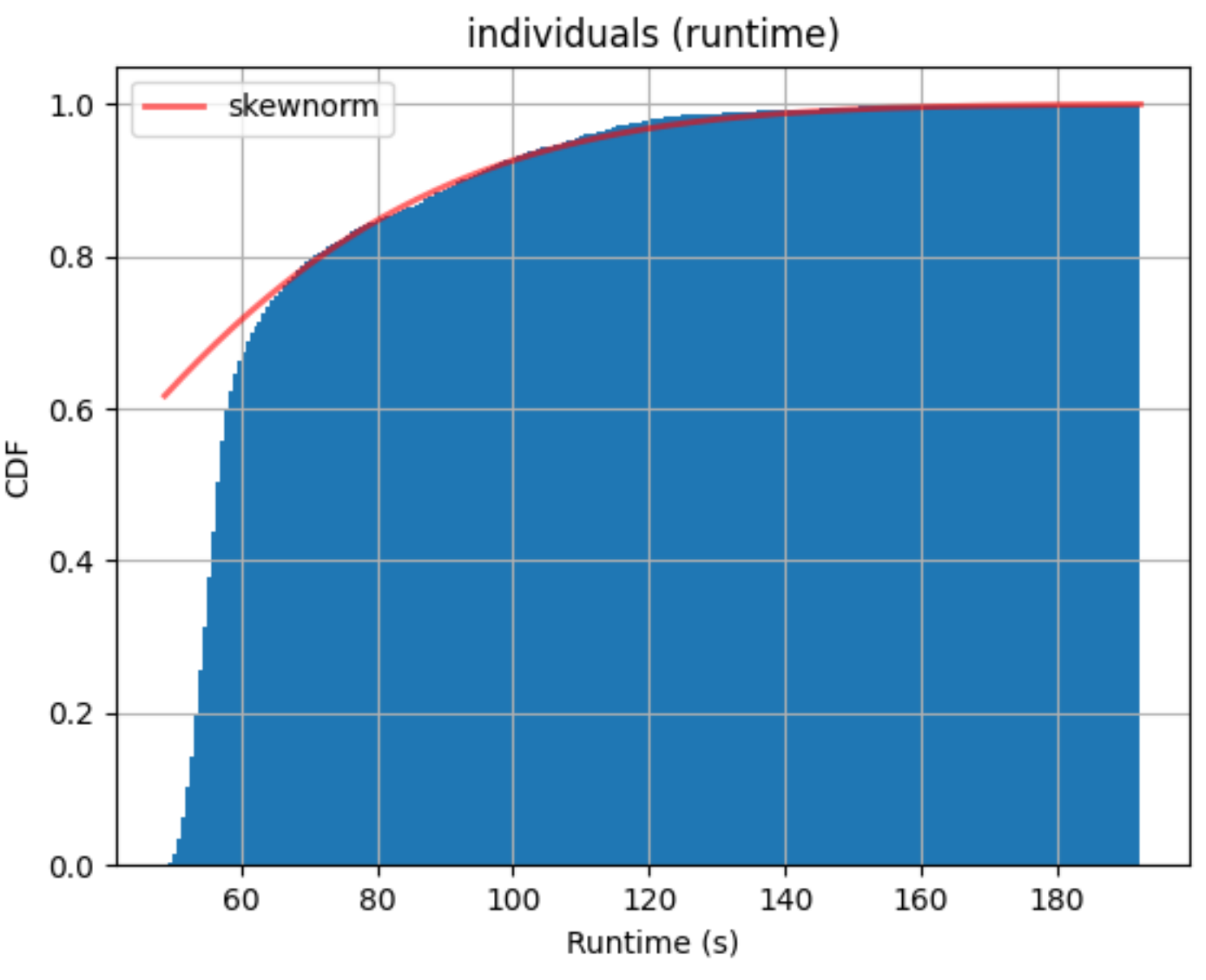} \\
  \includegraphics[width=0.7\linewidth]{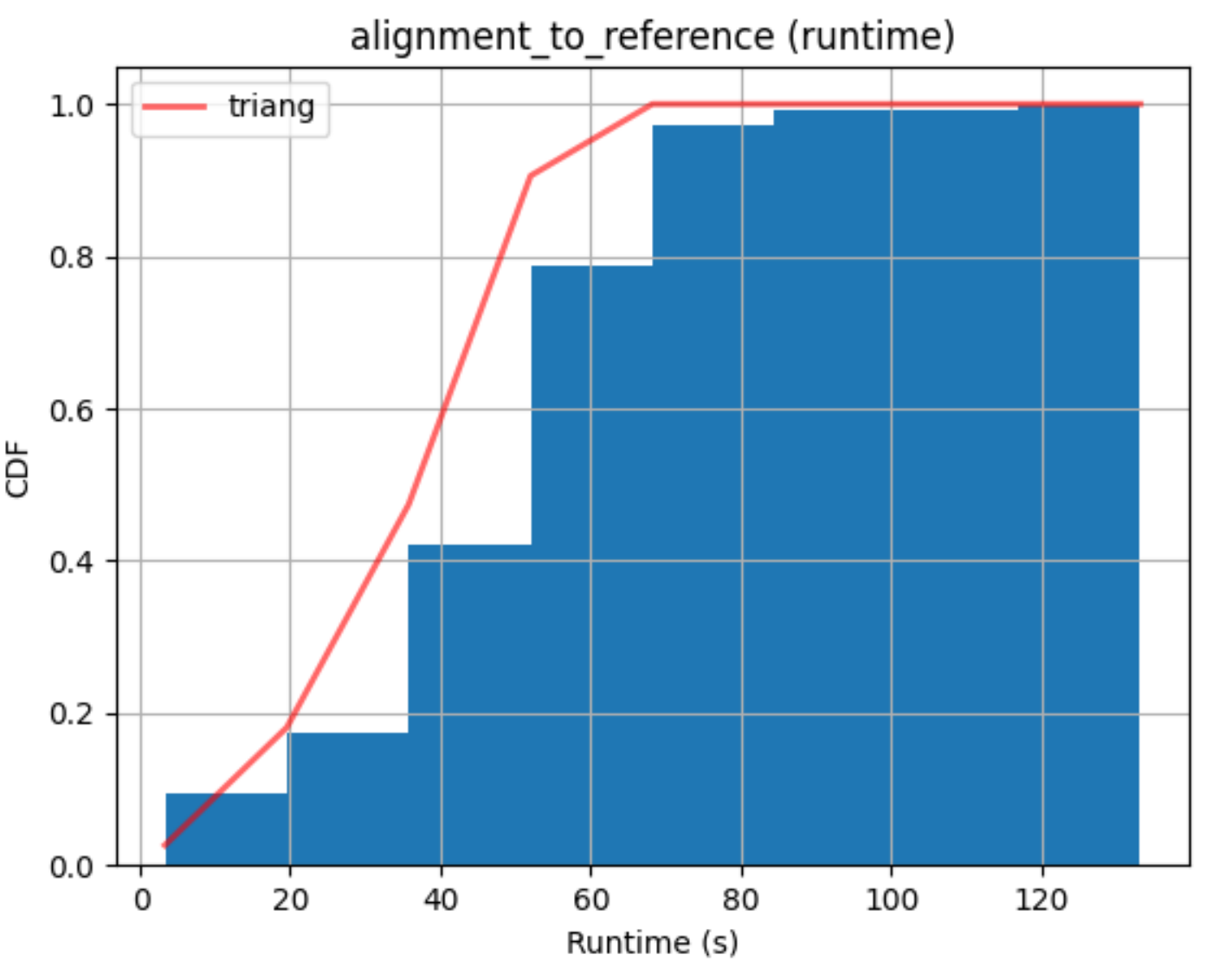}
  \caption{Example of probability distribution fitting of runtime
           (in seconds) for \texttt{individuals} tasks of the 1000Genome
           (\emph{top}) and \texttt{alignment\_to\_reference} tasks of the
           SoyKB (\emph{bottom}) workflows.}
  \label{fig:fitting-runtime}
\end{figure}

\pp{The WorkflowHub Python package}
In order to allow users to analyze existing workflow traces and to generate
synthetic workflow traces, the WorkflowHub framework provides a collection of
tools released as an open source Python package~\cite{workflowhub-python,
workflowhub-github}. This package provides several tools for analyzing
workflow traces. More specifically, analyses can be
performed to produce statistical summaries of workflow performance
characteristics, per task type. The package leverages the Python's
SciPy~\cite{virtanen2020scipy} package for performing probability
distributions fitting to a series of data to find the best (i.e., minimizes
the mean square error) probability distribution that represents the data.
In contrast to our previous work~\cite{ferreiradasilva-escience-2014},
which used only two probability distributions for generating workflow
performance metrics, the WorkflowHub's Python package attempts to fit data
with 23 probability distributions provided as part of SciPy's
statistics submodule. Figure~\ref{fig:fitting-runtime} shows an example of
probability distribution fitting of task runtimes for two task types from
different workflow traces, by plotting the cumulative distribution
function (CDF) of the data and the best probability distribution found.
The outcome of this analysis applied to an
entire workflow trace is a summary that includes, for each task type, the
best probability distribution fits for runtime, input data size, and output
data size. For instance, Table~\ref{tab:traces} lists, for each workflow
application for which WorkflowHub hosts traces, the probability
distributions used for these fits.  Listing~\ref{lst:recipe} shows the
summary (which is implemented as a Python object) for one particular task
type in the 1000Genome workflow application.  These summaries can then be
used to develop \emph{workflow recipes}, described hereafter.

\pp{Workflow Recipes}
The WorkflowHub Python package also provides a number of workflow
``recipes" for generating realistic synthetic workflow traces.  Each recipe
provides different methods for generating synthetic, yet realistic,
workflow traces depending on the properties that define the structure of
the actual workflow. A common method for generating synthetic traces
(regardless of the application) is to provide an upper bound for the total
number of tasks in the workflow. Although this functionality provides
flexibility and more control for generating an arbitrary number of
synthetic workflows, we have implemented mechanisms that define different
lower bound values (for each workflow recipe), so that the workflow
structure is guaranteed.  The workflow recipe also includes the summaries
obtained in the previous steps, so as to generate representative workflow
task instances (by sampling task runtimes and input/output data sizes using
the probability distribution in the summaries).  The current version of the
WorkflowHub's Python package provides recipes for generating synthetic
workflows for all 6 applications shown in Table~\ref{tab:traces}.  Detailed
documentation and examples can be found on the project's
website~\cite{workflowhub} and the online open access package
documentation~\cite{workflowhub-python}.

\definecolor{codegray}{rgb}{0.5,0.5,0.5}
\definecolor{codepurple}{rgb}{1,0,0}
\definecolor{backcolour}{rgb}{0.95,0.95,0.92}

\lstdefinestyle{mystyle}{
    backgroundcolor=\color{backcolour},
    numberstyle=\tiny\color{codegray},
    stringstyle=\color{codepurple},
    basicstyle=\ttfamily\scriptsize,
    breakatwhitespace=false,
    breaklines=false,
    captionpos=b,
    keepspaces=true,
    numbers=left,
    numbersep=1pt,
    showtabs=false,
    tabsize=2,
    columns=fullflexible,
    framexrightmargin=0em,
    float=tp,
    floatplacement=tbp,
}

\begin{lstlisting}[language=Python, style=mystyle, label={lst:recipe}, caption=Example of an analysis summary showing the best fit probability distribution for runtime of the \texttt{individuals} tasks (1000Genome workflow).]
"individuals": {
    "runtime": {
        "min": 48.846,
        "max": 192.232,
        "distribution": {
            "name": "skewnorm",
            "params": [
                11115267.652937062,
                -2.9628504044929433e-05,
                56.03957070238482
            ]
        }
    },
    ...
}
\end{lstlisting}

\subsection{Workflow Simulator}
\label{sec:simulator}

An alternative to conducting scientific workflow research via real-world
experiments is to use simulation.
Simulation is used in many computer
science domains and can address the limitations of real-world experiments.
In particular, real-world experiments are confined to those application
and platform configurations that are available to the researcher, and thus
typically can  only cover a small subset of the relevant scenarios that may be
encountered in practice. Furthermore, real-world experiments can be time-,
labor-, money-, and energy-intensive, as well as not perfectly
reproducible.

The third axis of the WorkflowHub project fosters the use of simulation for
scientific workflow research, e.g., the development of workflow scheduling
and resource provisioning algorithms, the development of workflow management
systems, and the evaluation of current an emerging computing platforms for
workflow executions.  We do not develop simulators as part of the
WorkflowHub project. Instead, we catalog open source workflow management
systems simulators (such as those developed using the WRENCH
framework~\cite{casanova-works-2018, casanova2020fgcs}) that support the
WorkflowHub common trace format. In other words, these simulators take as
input workflow traces (either from actual workflow executions, or
synthetically generated).  In the next section, we use one of the
simulators cataloged in the WorkflowHub project to quantify the extent
to which synthetic traces generated using WorkflowHub tools are representative
of real-world traces.

\section{Case Study: Evaluating Synthetic Traces with a Simulator of a Production WMS}
\label{sec:experiments}

In this section, we use a simulator~\cite{casanova2020fgcs} of a
state-of-the-art WMS, Pegasus~\cite{deelman-fgcs-2015}, as a case study for
evaluation and validation purposes. Pegasus is being used in production to
execute workflows for dozens of high-profile applications in a wide range
of scientific domains, and is the WMS we used to execute workflows on a
cloud environment for the purpose of collecting the traces described in
Section~\ref{sec:traces}.

The simulator is built using WRENCH~\cite{casanova-works-2018,
casanova2020fgcs}, a framework for implementing simulators of WMSs that are
accurate and can run scalably on a single computer, while requiring minimal
software development effort. In~\cite{casanova2020fgcs}, we have demonstrated
that WRENCH achieves these objectives, and provides high simulation accuracy
for workflow executions using Pegasus.

\subsection{Experimental Scenarios}

We consider experimental scenarios defined by particular workflow instances
to be executed on particular platforms. To assess the accuracy and
scalability of generated synthetic workflows, we have performed real workflow
executions with Pegasus and collected raw, time-stamped event traces from
these executions. These traces form the ground truth to which we can compare
simulated executions.

Actual workflow executions are conducted using the Chameleon Cloud platform, an
academic cloud testbed, on which we use homogeneous standard cloud units to
run an HTCondor pool with shared file system, a submit node (which runs Pegasus
and DAGMan), and a data node in the WAN. Each cloud unit consists of a
48-core 2.3GHz processor with 128 GiB of RAM. The bandwidth between the submit
node and worker nodes on these instances is about 10Gbps.

Whenever possible, for the experiments conducted in this section, we contrast
experiment results obtained with synthetic workflow traces generated with
WorkflowHub to results obtained using synthetic workflow traces generated using
our previous work~\cite{ferreiradasilva-escience-2014}.
In~\cite{casanova2020fgcs}, we have already demonstrated that the simulator
framework used in our previous work yields significant discrepancies from
actual executions. These discrepancies mostly stem from the use of a
simplistic network simulation model,
and from the simulator not capturing relevant details of the system, and
thus of the workflow execution.  Therefore, to reach fair conclusions regarding
the validity of synthetic workflow traces, in this paper we only use the more
accurate WRENCH simulator for all experiments.  Using this simulator we
quantify the extent to which each generated synthetic workflow trace (using
our previous work and using WorkflowHub) is representative of the original
real-world workflow trace.

The simulator code, details on the calibration procedure, and experimental
scenarios used in the rest of this section are all publicly available
online~\cite{wrench-pegasus}.

\subsection{Evaluating the Accuracy of Synthetic Traces}
\label{sec:accuracy}

To evaluate the accuracy of the generated synthetic workflow traces, we
consider 2 workflow applications: Montage and Epigenomics. We choose these
two applications to allow for comparison with our previous work -- the popular
generator in~\cite{ferreiradasilva-escience-2014}
can produce synthetic workflow traces for both these applications.  For each
real-world execution trace of each application, as archived on WorkflowHub,
we use WorkflowHub's Python package for generating a synthetic trace that
is similar to the actual execution trace (i.e., we bound the number of
tasks to the number of tasks in the actual workflow execution).  For
comparison purposes, we also generate synthetic traces using the generator
from our previous work.  To that end, we have calibrated that generator
with the parameter values shown in Table~\ref{tab:params}.

\begin{table*}[!t]
  \centering
  \setlength{\tabcolsep}{12pt}
  \small
  \begin{tabular}{lrrr}
    \toprule
    Application & \multicolumn{1}{c}{\# Tasks} & \multicolumn{1}{c}{Runtime Factor} & \multicolumn{1}{c}{Reference Size}  \\
    \midrule
    Epigenomics & [125, 263, 405, 559, 713, 803] & 0.001 & [0.1, 16384] \\
    Montage & [1738, 4846, 7117, 9805] & 0.05 & [0.1, 113774] \\
    \bottomrule
  \end{tabular}
  \caption{Parameter values used for calibrating the Workflow Generator from our previous work~\cite{ferreiradasilva-escience-2014}.}
  \label{tab:params}
\end{table*}

\begin{figure}[!t]
  \centering
  \includegraphics[width=\linewidth]{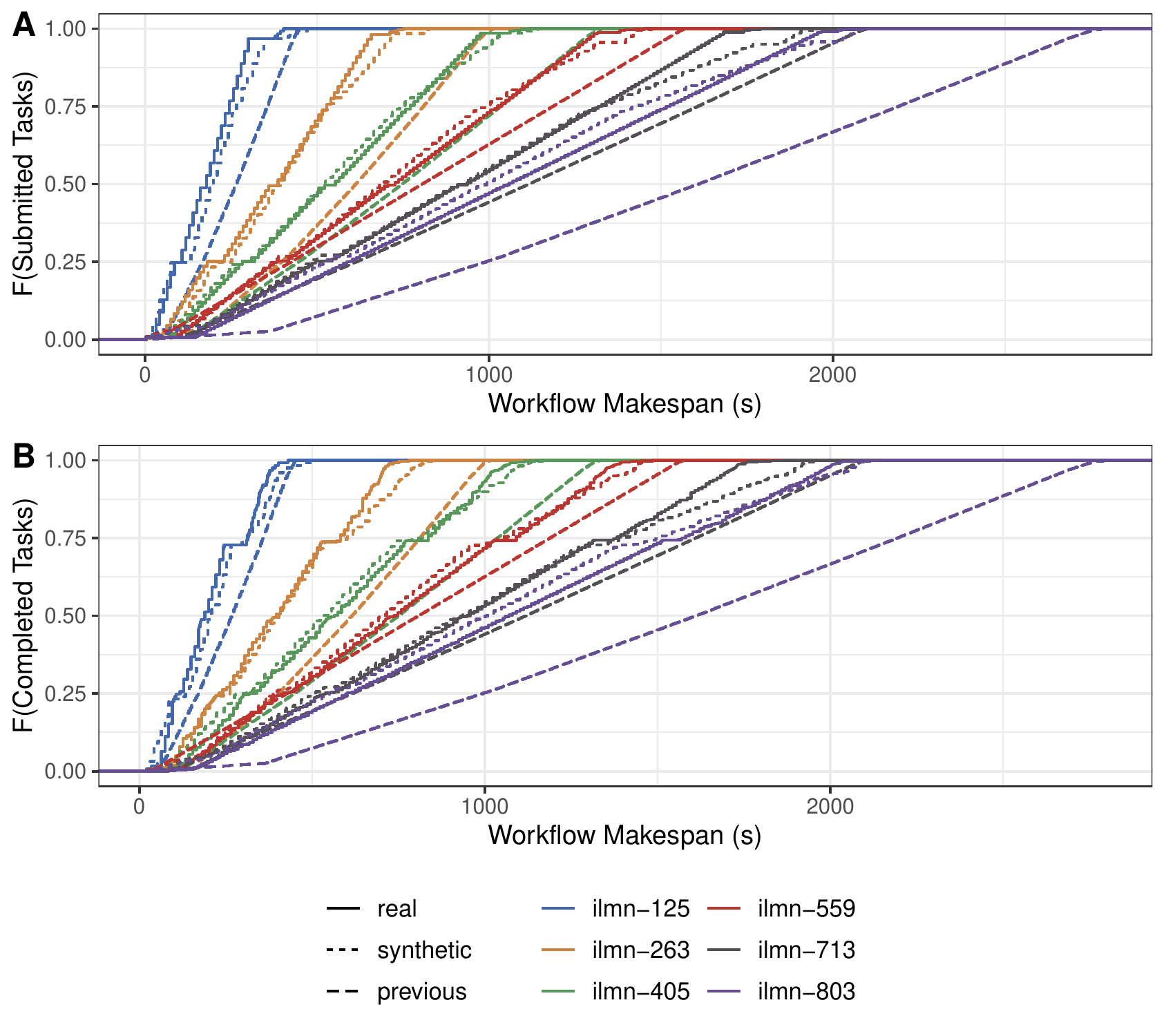}
  \vspace{-20pt}
  \caption{Empirical cumulative distribution function of task submit times
           (top) and task completion times (bottom) for sample real-world
           (``real") and synthetic (``synthetic" and ``previous") workflow
           trace executions of Epigenomics using the WRENCH-Pegasus simulator.}
  \label{fig:ilmn}
\end{figure}

\pp{Epigenomics}
Figure~\ref{fig:ilmn} shows simulated empirical cumulative distribution functions
(ECDFs) of task submission dates (top) and task completion dates (bottom), for sample runs of real-world and
synthetic workflow trace executions of the Epigenomics workflow used on the ILMN dataset.
We observe that WorkflowHub's generated synthetic
workflow traces (``synthetic") yields very similar simulated execution
behavior when compared to the real-world execution trace (``real") -- the
averaged root mean squared error (RMSE) is 49.02 for task submission
and 50.90 for task completion. Recall that small discrepancies in the workflow
execution behavior are expected since the workflow task
characteristics are sampled from probability distributions. For the synthetic
traces generated with the previous generator (``previous"), the averaged
RMSEs are 251.87 for task submission and 224.15 for task completion.
These substantial discrepancies in execution behavior are mostly due to lack
of using accurate probability distributions to model task runtimes, and
input and output file sizes.  While WorkflowHub's generator produces
synthetic traces using the best fitted distribution per workflow task type
(as described in~\Cref{sec:traces}), the previous generator only uses
uniform and truncated normal distributions.  Moreover, file size generation in that
generator uses a random generation process that leads to inconsistent
distribution of file sizes (i.e., the initial reference size seed may shift
the density of the distribution toward a lower density area when compared
to the empirical distributions obtained from actual workflow executions).

\begin{figure}[!t]
  \centering
  \includegraphics[width=\linewidth]{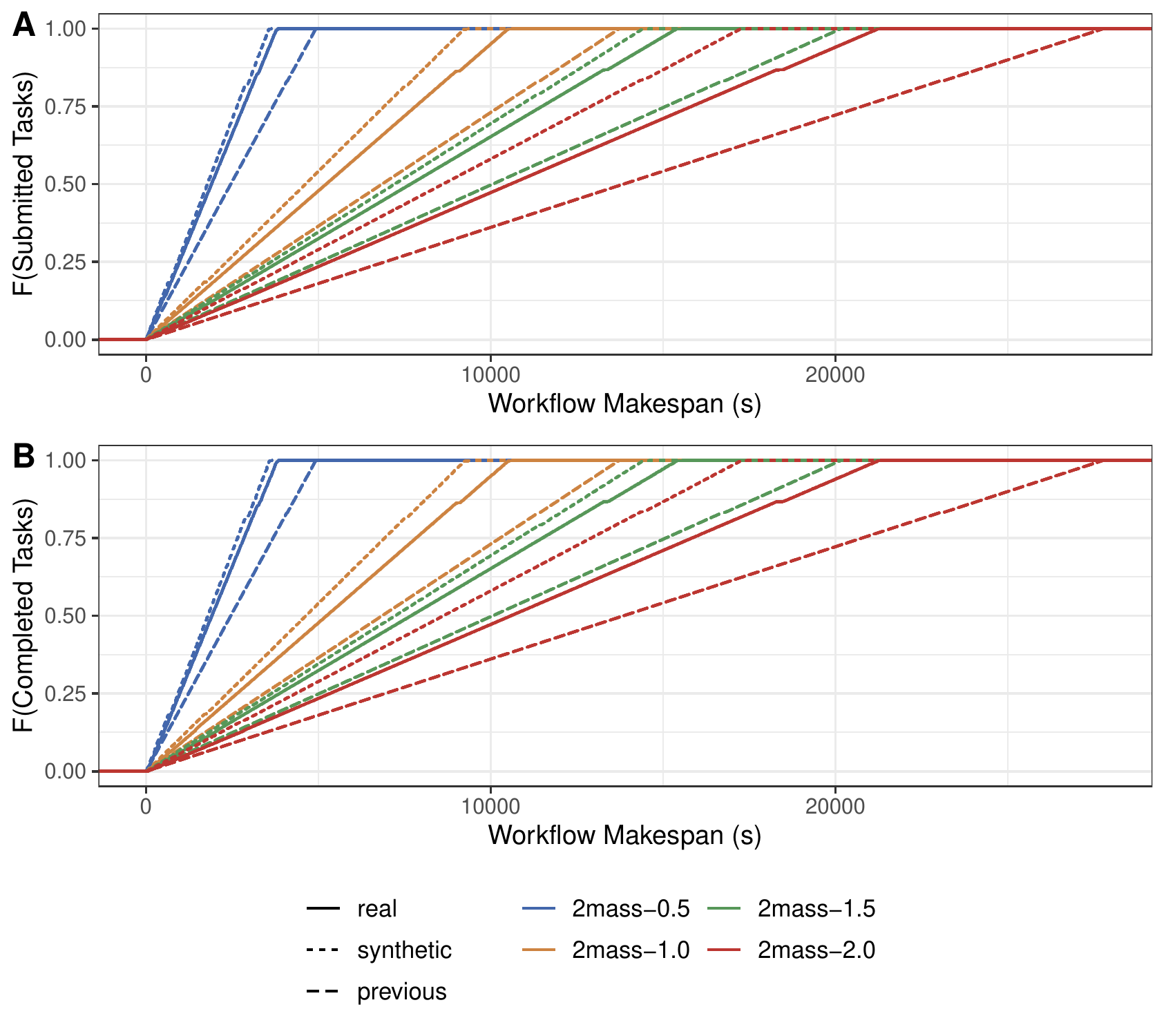}
  \vspace{-20pt}
  \caption{Empirical cumulative distribution function of task submit times
           (top) and task completion times (bottom) for sample real-world
           (``real") and synthetic (``synthetic" and ``previous") workflow
           trace executions of Montage using the WRENCH-Pegasus simulator.}
  \label{fig:2mass}
\end{figure}

\pp{Montage}
Figure~\ref{fig:2mass} shows simulated ECDFs for sample runs of real-world and
synthetic workflow trace executions of Montage for the 2MASS dataset. Similarly
to the Epigenomics results above, WorkflowHub's generated synthetic workflow traces
(``synthetic") produce workflow execution behaviors close to that of the real-world
execution trace (``real") -- average RMSEs are 39.82 for task submission and
46.93 for task completion. When contrasted to synthetic traces generated with
the preceding generator (``previous''), RMSEs are 2265.73 and 2253.54 for
task submission and completion, respectively. These wide discrepancies are
due to the larger number of tasks in the workflow, and Montage's idiosyncratic
workflow structure (extreme fan-in/out pattern). In our previous generator, the
generation of file sizes, in particular the larger ones, follows a truncated
normal distribution, in which the large variance shortens and broadens the
curve, thus generating unbalanced data sizes. As a result, the effect
is that tasks in the workflow's critical path are delayed by data movement operations, causing the workflow makespan increases. This effect is an artifact
of the synthetic trace, an is not seen in real-world executions.

Overall, when using the generated synthetic traces in simulation, we
observe large discrepancies when these traces were generated using our
previous work. By contrast, using WorkflowHub's generation methods
produces workflow traces that closely match the real-world execution
behaviors (even though some discrepancies necessarily remain due to random
sampling effects).

\subsection{Evaluating the Scaling of Synthetic Traces}

In this section, we evaluate the accuracy of the structure of synthetic
workflow traces generated based on collected traces at a lower scale, where
the scale is the number of workflow tasks. In other words, when generating
synthetic workflow traces at various scales, we want to see whether the
overall workflow structure is still representative of the workflow application.

We perform experiments using the WRENCH-Pegasus simulator for each workflow
application supported by the WorkflowHub project. For each application, we
run the simulator for a reference workflow trace (from a real-world
execution), and for synthetic traces in which the upper bound limit for the
number of tasks are 1K, 5K, 10K, 25K, 50K, and 100K. The goal is to determine
whether the simulated execution pattern using these synthetic traces
are consistent with that observed when using the reference trace.

\begin{figure}[!t]
  \centering
  \includegraphics[width=\linewidth]{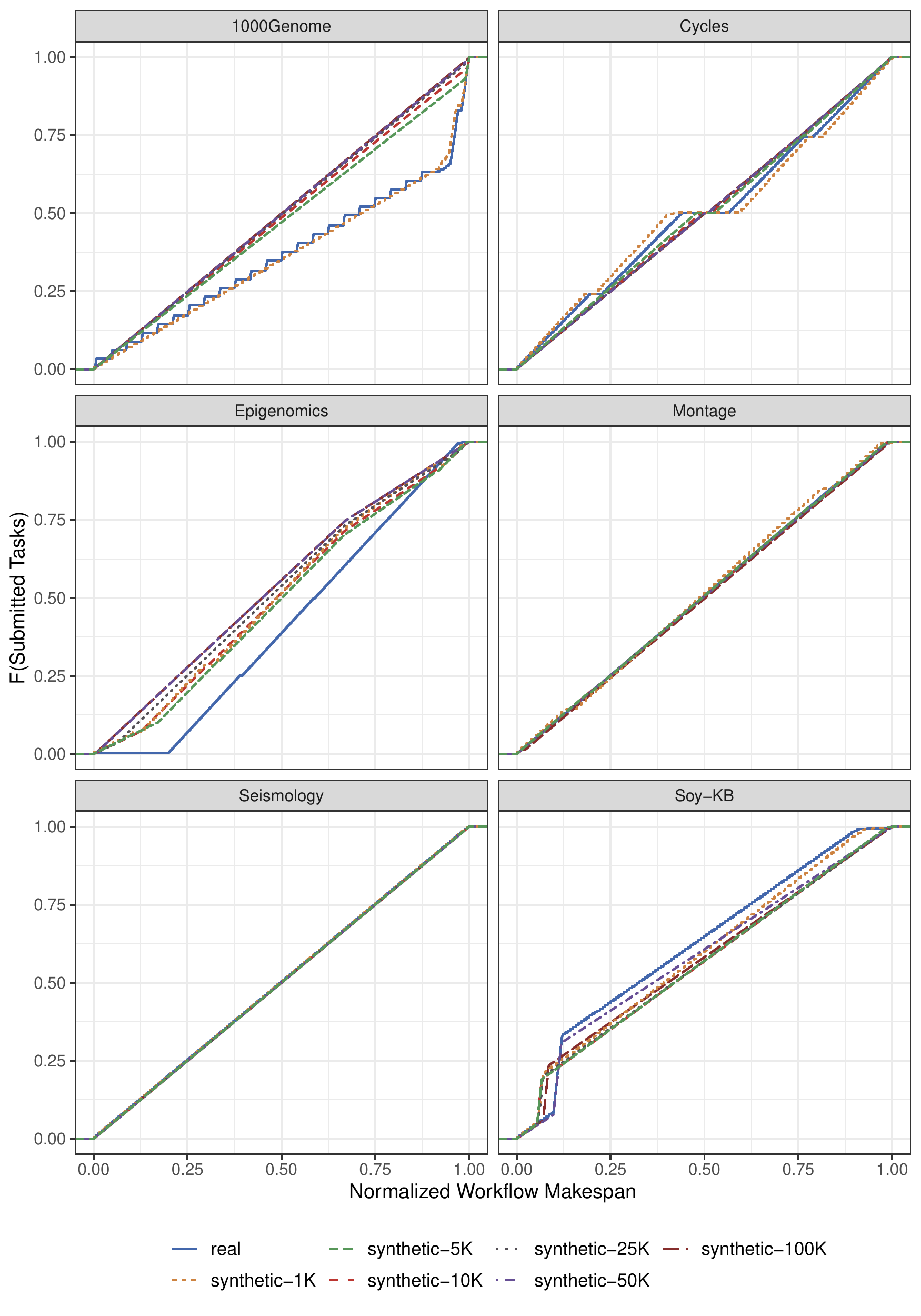}
  \caption{Empirical cumulative distribution function of task submit times
           for sample real-world (``real") and synthetic (``synthetic-\emph{n}K",
           $n~\in~[1, 5, 10, 25, 50, 100]$)
           workflow trace executions using the WRENCH-Pegasus simulator.
           Workflow makespan has been normalized for comparison purposes.}
  \label{fig:scalability}
\end{figure}

Figure~\ref{fig:scalability} shows ECDFs for task submit times and task
completion times for sample runs of these configurations. This figure
shows \emph{normalized} makespan values on the horizontal axis. This is
to make it possible to use visual inspection for assessing
whether the structure of the generated workflows conform with the
reference workflow, even though the executions of these workflows have
different
makespans.  In addition, \Cref{tab:scale} shows the number of tasks that
compose the reference workflow trace for each application (``real'') and
the RMSE values for the synthetic workflows (``synthetic-\emph{n}K") when
contrasted to the reference workflow execution trace.

Overall, generated synthetic traces lead to fairly similar execution
patterns when compared to the reference trace. Not surprisingly, the
Seismology workflow, due to its simple structure, shows ideal scalability behavior
(RMSE values are nearly zero as observed in \Cref{tab:scale}) -- the
workflow structure follows a simple \emph{merge} pattern, in which a set of
tasks for computing seismogram deconvolutions are followed by a single
task that combines all processed fits. Small, yet expected, discrepancies
are observed for the Cycles, Montage, and SoyKB workflows. As discussed in
\Cref{sec:accuracy}, these discrepancies are due to the random generation of
workflow task characteristics, which are drawn from various probability
distributions.

\begin{table*}[!t]
  \centering
  \setlength{\tabcolsep}{12pt}
  \small
  \begin{tabular}{lrrrrrrr}
    \toprule
    \multirow{2}{*}{Application} & \multicolumn{1}{c}{\# Tasks} & \multicolumn{6}{c}{RMSE} \\ \cline{3-8}
    & (real) & 1K & 5K & 10K & 25K & 50K & 100K \\
    \midrule
    1000Genome  & 903 & 0.03 & 0.60 & 0.66 & 0.69 & 0.70 & 0.71 \\
    Cycles      & 331 & 0.05 & 0.18 & 0.26 & 0.27 & 0.34 & 0.36 \\
    Epigenomics & 125 & 0.30 & 0.53 & 0.49 & 0.56 & 0.61 & 0.61 \\
                & 1095 & 0.05 & 0.06 & 0.06 & 0.06 & 0.07 & 0.07 \\
    Montage     & 1738 & 0.25 & 0.35 & 0.16 & 0.22 & 0.41 & 0.48 \\
    Seismology  & 101 & 0.05 & 0.04 & 0.05 & 0.05 & 0.05 & 0.05 \\
    SoyKB       & 383 & 0.07 & 0.22 & 0.28 & 0.37 & 0.37 & 0.38 \\
    \bottomrule
  \end{tabular}
  \caption{Root mean square errors (RMSE) for large scale synthetic workflows. (RMSE values are computed from normalized workflow makespan.)}
  \label{tab:scale}
\end{table*}

Visual inspection of Figure~\ref{fig:scalability} indicates a
significant divergence in execution behaviors for the 1000Genome and
Epigenomics workflows.  For the 1000Genome workflow, the difference between
execution behaviors of the reference and \emph{synthetic-1K} workflows is
minimum, which is expected as the reference workflow is composed of 903
tasks, which is close to the 1,000 tasks in the synthetic workflow (see
\Cref{tab:scale}).  However, for synthetic workflows with 5K tasks and
higher, the execution pattern does not conform to that with the reference
trace (and RMSE values indicate large errors). After carefully inspecting
the ECDFs and the workflow structure representation, we noticed that the
structure of the 1000Genome workflow, at is scales up, leads to an
increase on the number of
tasks in the upper levels of the workflow (which are pipelines composed of
6 tasks each), followed by a few tasks that combine the results of these
tasks (seen in Figure~\ref{fig:scalability} as the spike around 90\% of the
workflow execution time). When increasing the number of tasks for the
1000Genome workflow, this ``spike'' becomes smoother and fades out in the
distribution. A similar structural characteristic is also observed for the
Epigenomics workflow. To evaluate this hypothesis, we have also performed
runs with an execution trace for the Epigenomics workflow composed of 1095
tasks (also shown in \Cref{tab:scale}, but not shown in
Figure~\ref{fig:scalability}), which produces very similar execution
behaviors when compared to the synthetic workflows. We were not able to
replicate this experiment for the 1000Genome workflow as we are unable to
run the actual application with more than 903 tasks.

Overall, our experiments results show that the tools provided as part of the
WorkflowHub framework not only can be used to generate representative
synthetic workflow traces (i.e., with workflow structures and task
characteristics distributions that resembles those in traces obtained from
real-world workflow executions), but can also generate representative
workflow traces at larger scales that of available workflow traces.
This is crucial for supporting ongoing research that targets large-scale
executions of complex scientific applications on emerging platforms.

\section{Conclusion}
\label{sec:conclusion}

In this paper, we have presented the WorkflowHub project, a community
framework for archiving workflow execution traces, analyzing these traces,
producing realistic synthetic workflow traces, and simulating workflow
executions using all these traces. WorkflowHub provides a collection of
resources for developing workflow recipes based on traces collected from
the execution of real-world workflow applications.  These workflow recipes
are then used to produce synthetic, yet realistic, workflow traces that can
enable a variety of novel workflow systems research and development
activities. Via a case study using an accurate and scalable simulator of a
production WMS, we have demonstrated that WorkflowHub achieves these
objectives, and that it favorably compares to a widely used previously
developed workflow generator tool. The main finding is that, with
WorkflowHub, one can generate representative synthetic workflow traces at
various scales in a way it preserves the workflow application's key features.
WorkflowHub is open source and welcomes contributors. It currently provides
a collection of 101 traces from actual workflow executions, and can
generate synthetic workflows from 6 applications from 4 science domains.
Version 0.3 was released in August 2020. We refer the reader to
\url{https://workflowhub.org} for software, documentation, and links to
collections of traces and simulators.

A short-term development direction is to use statistical learning methods,
such as regression analysis, for automating the process of generating
workflow recipes -- specifically the description of the workflow structure,
i.e., relations between tasks and dependencies. We also intend to provide
continuous supported development of novel workflow recipes to broaden the
number of science domains in which WorkflowHub can potentially impact
research and development efforts. Finally, another future direction is
to use synthetic workflows to support the development of simulation-driven
pedagogical modules~\cite{tanaka2019eduhpc}, which include targeted activities
through which students acquire knowledge by experimenting with various
application and platform scenarios in simulation.

\bibliographystyle{abbrv}
\bibliography{references}

\balance

\end{document}